\documentclass[a4paper]{VisionStyle}
\usepackage{epsfig}

\begin{document}

\title{On the Temperature and Intensity Distribution of the Galactic X-ray Plasma}

\author{M.\,Kappes \and J.\,Pradas \and J.\,Kerp} 

\institute{Radioastronomisches Institut der Universit\"at Bonn, Auf dem H\"ugel 71, 53121 Bonn, Germany}

\maketitle 

\begin{abstract}
We present the results of our investigation of the composition of the diffuse soft X--ray 
background emission (SXRB). Combining data of the Leiden/Dwingeloo {\sc HI} Survey and
the {\it ROSAT} All--Sky Survey (RASS), we set up a radiation transport equation in order to
model the SXRB. Two different techniques lead to the model parameters: An image 
oriented approach which compares observed and modeled maps of the 1/4 and 3/4\,keV X--ray energy 
regime and a more analytic approach using scatter diagrams.

The analysis shows that {\em only three} independent components of the emitting plasma 
(local, halo and extragalactic) are needed to explain the SXRB. The results 
for the temperatures and X--ray intensities, which characterize the three components, are given 
and compared to an alternative model.

\end{abstract}

\section{Introduction}
To understand the origin and evolution of the Milky Way it is necessary 
to investigate its emission across the entire electromagnetic frequency
spectrum. In the soft X--ray energy regime much progress was gained by the
{\it ROSAT}--mission and its discovery of coronal gas located within the Milky Way halo
(\cite{jkerp-E1-6:sno91}).

Today there is general agreement that we can identify at least three
individual components which contribute to the diffuse soft X--ray background
(SXRB) emission: The coronal gas partly filling the Local Bubble (\cite{jkerp-E1-6:sno98}), X--ray plasma
localized within the Milky Way halo (\cite{jkerp-E1-6:pie98}) and finally the superposed emission of 
individual X--ray sources at extragalactic distances (\cite{jkerp-E1-6:has01}).

With {\it ROSAT}--all sky survey data it is possible to shed light on the
physical properties of the several components of the SXRB. Questions we like to answer are: Is the 
halo plasma hotter or cooler than the local X--ray gas? Is there evidence for more than one 
single coronal gas phase in the Milky Way halo (\cite{jkerp-E1-6:kun00})? Is the plasma 
emissivity a function of Galactic longitude and/or latitude (\cite{jkerp-E1-6:pie98})?

\section{Data and Model}
The correlation of the {\it ROSAT} All--Sky Survey (RASS) 1/4 keV ({\it ROSAT}--C--band) 
and 3/4 keV ({\it ROSAT}--M--band) data with the Leiden/Dwingeloo {\sc HI} Survey of 
galactic neutral hydrogen (\cite{jkerp-E1-6:har97}) provides an opportunity to disentangle 
the different SXRB components. Figure \ref{jkerp-E1-6:rad} illustrates our approach to model 
the X--ray radiation transport through the Galactic interstellar medium. Because of the anti--correlation 
between X--ray radiation and {\sc HI} column density it is possible to set up the following 
radiation transport equation:

\begin{equation}
I = I_{\rm l} + I_{\rm h} \cdot e^{- \sigma(E, N_{\rm HI,h}) \cdot N_{\rm HI,h}}+ I_{\rm e} \cdot e^{- \sigma(E, N_{\rm HI,e}) \cdot N_{\rm HI,e}}
\label{jkerp-E1-6:rad}
\end{equation}

\noindent
$I_{\rm l}$ denotes the Local Bubble component, $I_{\rm h}$ denotes the halo component
and $I_{\rm e}$ the extragalactic contribution. The observed X--ray intensity distribution 
is modulated by photoelectric absorption traced by the {\sc HI} gas. In our initial approach 
we include three X--ray emission 
components. First, an unabsorbed foreground Raymond--Smith plasma representing the Local 
Bubble emission, an absorbed distant Raymond--Smith plasma (halo component) and
an absorbed extragalactic energy power--law (EPL) with index $\alpha = -1.5$ (\cite{jkerp-E1-6:has01}). 

\begin{figure}[!ht]
\begin{center}
\psfig{file=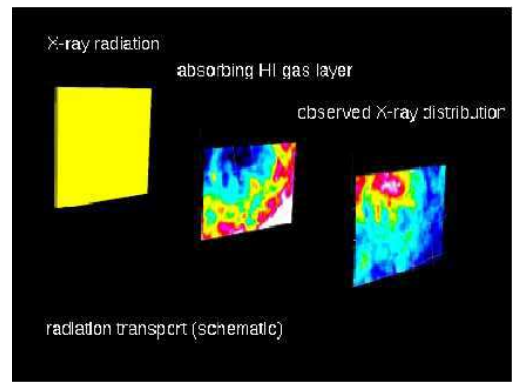, width=5cm, angle=270}
\end{center}
\caption{A homogeneous X--ray background is absorbed by the Galactic {\sc HI}--layer and produces the observed X--ray distribution. This is just a simple illustration of the radiation transport approach according
to Eq. 1}  
\label{jkerp-E1-6:rad}
\end{figure}

\section{Deriving the model parameters}
\begin{figure}[!ht]
\begin{center}
\psfig{file=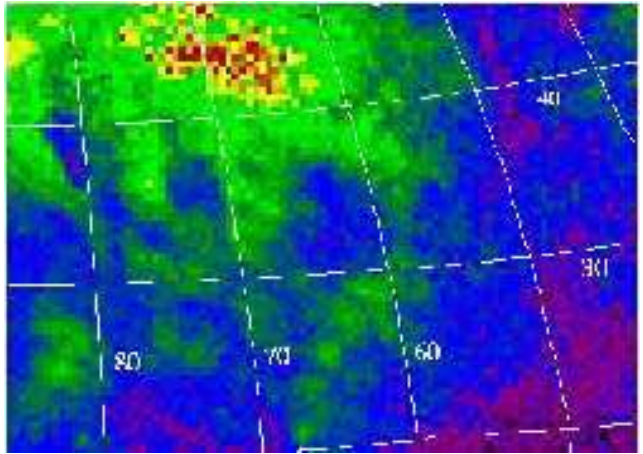, width=6cm, angle=270}
\end{center}
\caption{Observed X--ray intensity of {\bf field A}, {\it ROSAT}--C--band.}
\label{jkerp-E1-6:loco}
\end{figure}

We investigate two different fields: {\em Field A} at 
$20{\degr} < b < 47{\degr}$, $34{\degr} < l < 85{\degr}$ and {\em field B} at 
$12{\degr} < b < 74{\degr}$, $99{\degr} < l < 166{\degr}$. These fields are at high Galactic 
latitude, where the smaller {\sc HI} column density allows a better study of the halo component in 
comparison with the Galactic plane where this radiation is much stronger absorbed. Both fields cover
a large range in {\sc HI} column densities which improves the significance of the X--ray/{\sc HI} 
correlation. Figure \ref{jkerp-E1-6:loco} shows the 1/4 keV {\it ROSAT}--band for {\it field A}.

\subsection{Scatter diagrams}
\begin{figure}[!ht]
\begin{center}
\psfig{file=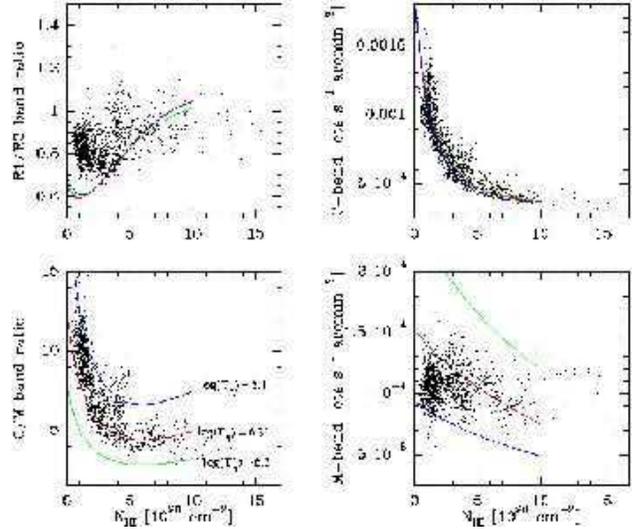, width=7cm, angle=270}
\end{center}
\caption{Scatter diagrams and energy band--ratios for {\bf field B}. The three curves correspond to
three different halo temperatures. Only the M--band and the C/M--band ratios are a good
indicator to the different temperature estimates in contrast to the C--band. Note, that the C--band
scatter diagram is not sensitive for different temperatures, i.e. the three curves are almost the same
in the upper right diagram.}
\label{jkerp-E1-6:losca}
\end{figure}

First, we evaluate the X--ray intensity of the Local Bubble. For this aim, we 
produce ``scatter diagrams'' which are shown in Fig. \ref{jkerp-E1-6:losca}. 
At high column densities the distant and extragalactic 
X--ray components are so strongly absorbed that the remaining C--band intensity can be attributed 
entirely to the Local Bubble emission. We derive a Local Bubble intensity of 
$I_{\rm l} = 350 \cdot 10^{-6} {\rm cts \enspace s^{-1}\,arcmin^{-2}}$ for both fields investigated.

Second, for the extragalactic background intensity we use the value  $I_{\rm e} = (228 \pm 90) \cdot 10^{-6} {\rm cts \enspace s^{-1}\,arcmin^{-2}}$ given by \cite*{jkerp-E1-6:bar96}, as a first estimate. 
The power--law index is fixed to $\alpha = -1.5$ (\cite{jkerp-E1-6:has01}).

Third, we evaluate the contribution of the halo component to the SXRB. Different values for the 
C--band intensity and temperatures of the Raymond--Smith plasma used, is combined with the 
corresponding theoretical band ratios.

\begin{figure}[!ht]
\begin{center}
\psfig{file=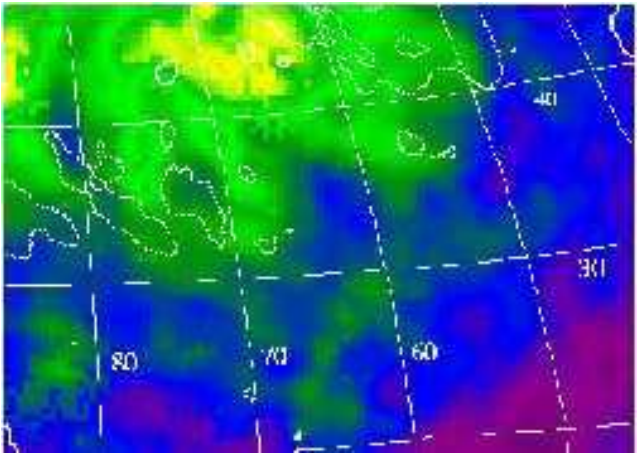, width=6cm, angle=270}
\end{center}
\caption{Modeled X--ray intensity of {\bf field A}, {\it ROSAT}--C--band. The contours show areas for which the
deviations between model and observation are greater than 2.5$\sigma$.}  
\label{jkerp-E1-6:locm}
\end{figure}

It turns out that the C--band scatter diagram is {\em not} a sensible measure for halo plasma temperature; 
on the contrary the other energy bands and derived ratios are! Note especially the upper right diagram
in Fig. \ref{jkerp-E1-6:losca}: The curves for the three different temperatures are almost identical. 
Only in combination with the M--band scatter diagram and the C/M--band ratio it is possible to
derive the temperatures reliably.
Because of this finding, we have to fit {\em simultaneously} the R1--, R2--, C-- and M--band data. 
The derived results -- on temperature and intensity -- are used to calculate model images 
(see Fig. \ref{jkerp-E1-6:locm}) which we test independently as follows.

\begin{figure}[!ht]
\begin{center}
\psfig{file=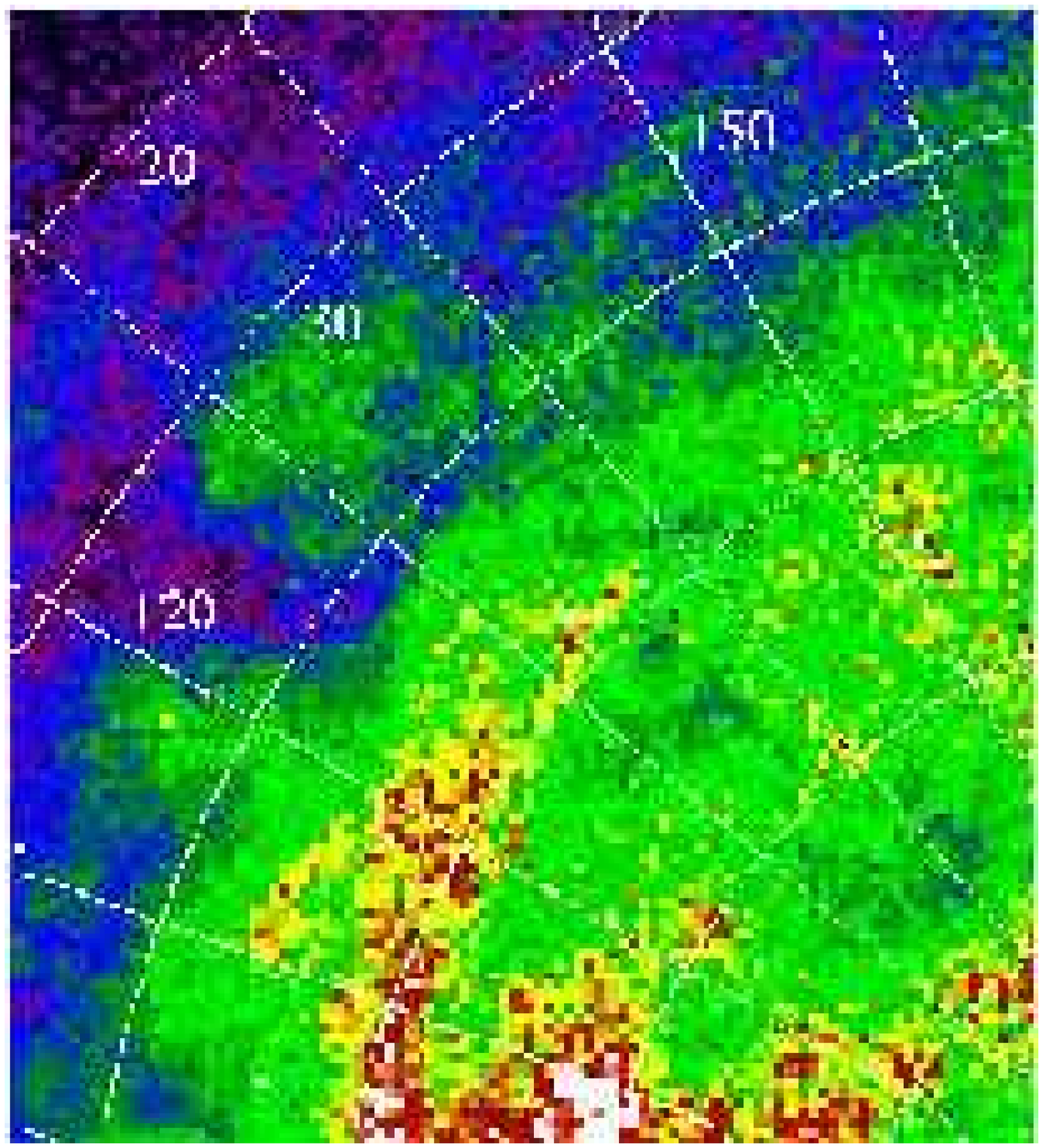, width=6cm, angle=270}
\end{center}
\caption{Observed X--ray intensity of {\bf field B}, {\it ROSAT}--C--band.}
\label{jkerp-E1-6:hico}
\end{figure}

\begin{figure}[!ht]
\begin{center}
\psfig{file=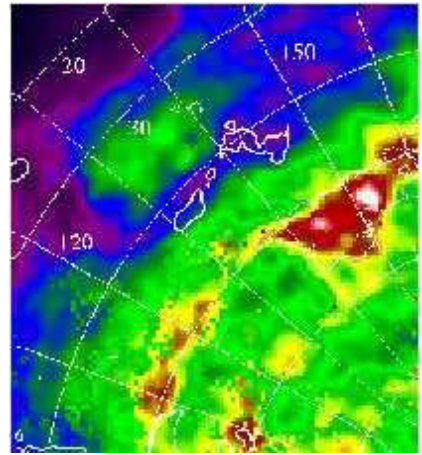, width=6cm, angle=270}
\end{center}
\caption{Modeled X--ray intensity of {\bf field B}, {\it ROSAT}--C--band. The contours show areas for which the
deviations between model and observation are greater than 2.5$\sigma$. The dotted contours towards the
Lockman--Area ($(l,b)=(150{\degr}, 53{\degr})$) indicate a lack of absorbing material, while the solid
encircled area at $(l,b)=(140{\degr}, 40{\degr})$ corresponds to the M81 group of galaxies which 
increases the Galactic {\sc HI} column density. Therefore the modeled X--ray intensity is too faint.}  
\label{jkerp-E1-6:hicm}
\end{figure}

\subsection{Modeled images}
By subtracting the modeled images from the observed ones (this is done pixel--by--pixel) 
we calculate absolute deviations 
between both which are divided by the uncertainty maps of the {\it ROSAT} observations in both 
C-- and M--band. The resulting normalized deviation--distribution is a measure for the 
statistical significance of the model expressed in units of the standard deviation 
(e.g. if the deviation width is close to unity then the uncertainty 
of the model is comparable to the uncertainty of the {\it ROSAT} data).

We produce ``deviation images'' and superimpose their contour lines on the {\it ROSAT}--C--band model 
images, as shown in Fig. \ref{jkerp-E1-6:locm}. At first glance it is clear that the 
deviations are not randomly distributed across the fields, but form coherent structures. 
The dotted contours correspond to areas where the modeled intensities are brighter than the 
observed ones, while the modeled intensity encircled by the solid contours is too faint. 
The regions with dotted contours can be explained by the lack of absorbing material, 
while the regions marked by solid lines represent excess emission or too much absorbing material. 
For example, {\it field B} contains intergroup {\sc HI}--gas belonging the M81 group of galaxies, 
which yields X--ray excess emission in the modeled intensity (see Fig. \ref{jkerp-E1-6:hicm}). 

In order to minimize the misfitting regions, we 
subtract gas with velocities $|{\rm v}_{\rm LSR}| \ge 25\,{\rm km\,s^{-1}}$ 
from the {\sc HI} data in the solid marked areas and it turns out that the model fits much better, 
without producing further excess emission. With this {\sc HI} data selection, which is done
for both fields, it is possible to reduce a huge area of excess emission in 
Fig. \ref{jkerp-E1-6:locm} to only three small spots. Unfortunately the dotted features 
cannot be identified. 

\section{Testing the model}
\begin{figure}[!ht]
\begin{center}
\psfig{file=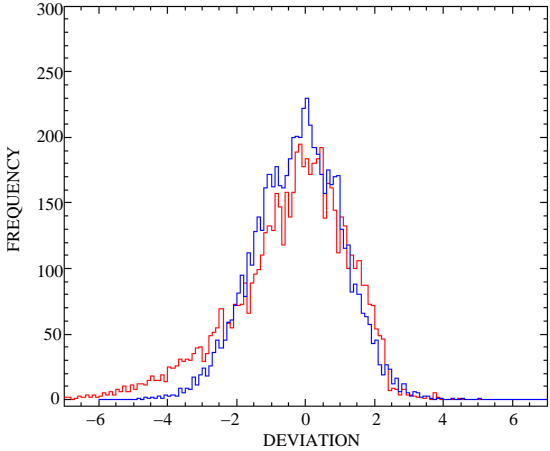, width=6cm, angle=270}
\end{center}
\caption{This histogram refers to field A where the red curve is the deviation of the C--band and the blue curve corresponds to the M--band.}  
\label{jkerp-E1-6:hist}
\end{figure}

To obtain better model parameters we take a closer look at the statistics of the deviations. 
Figure \ref{jkerp-E1-6:hist} illustrates the goodness of the model. The red plot represents 
the deviation in the C--band while the blue plot corresponds to the M--band deviation.

\begin{figure}[!ht]
\begin{center}
\psfig{file=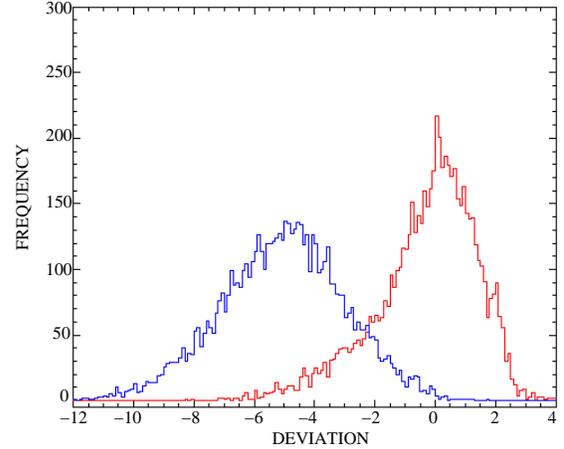, width=6cm, angle=270}
\end{center}
\caption{Statistical significance for the two distant--component model by Kuntz \& Snowden (2000) for field A. The red curve is the deviation of the C--band and the blue curve corresponds to the M--band.}  
\label{jkerp-E1-6:histsno}
\end{figure}

The goal is to derive model parameters which lead to histograms with a mean $\mu = 0$ and a standard 
deviation $\sigma = 1$ simultaneously in both energy bands. This is done by an iterative process 
in which we vary the model parameters we initially obtained from the scatter diagrams and 
analyze the mean and standard deviation of their histograms. Varying $I_{\rm e}$, it turns out,
that a value of $170 \cdot 10^{-6} {\rm cts \enspace s^{-1}\,arcmin^{-2}}$ fits much better 
than the original value given by \cite*{jkerp-E1-6:bar96}. 
In addition to our model we fit another model with two distant halo components as proposed 
by \cite*{jkerp-E1-6:kun00}. For the statistical significance see Fig. \ref{jkerp-E1-6:histsno}.
Note, that both energy bands are independent from each other in the two distant halo component model, 
i.e. they cannot be fitted simultaneously, in contrast to our approach.

The best fitting model intensities are shown in Tab. \ref{jkerp-E1-6:result} for the 
observed fields. The best fitting temperature for the Local Bubble is log($T_{\rm l}$)=5.9 and 
for the Halo we derive log($T_{\rm h}$)=6.2\,. For the two halo component model we adopt
the temperatures derived by \cite*{jkerp-E1-6:kun00}: log($T_{\rm l}$)=6.1, log($T_{\rm h_1}$)=6.0, and
log($T_{\rm h_2}$)=6.4\,. 

\begin{table}[!ht]
  \begin{center}
    \leavevmode
    \footnotesize
     \begin{tabular}[h]{ccccc}
        {\bf Field A}   & $I_{\rm l}$   & $I_{\rm h}$   & $I_{\rm e}$ \\
        \hline \\[-8pt]
                1/4 keV & 350   & 1975  & 170   \\
                3/4 keV & 1     & 152   & 40    \\
        \hline \\[-8pt]
        {\bf Field B}           &       &       &       \\
        \hline \\[-8pt]
                1/4 keV & 350   & 1380  & 170   \\
                3/4 keV & 1     & 110   & 40    \\
        \hline \\[-8pt]
        {\bf Field A}           &       &       &       \\
        \hline \\[-8pt]
                1/4 keV & 350   & 1380,1100& 230\\
                3/4 keV & 12    & 15,282& 53    \\
        \hline \\[-8pt]
     \end{tabular}
  \caption{The upper two sections compile the best fitting intensities for the fields A and B. The third section contains the values for field A according to the two halo component model by Kuntz \& Snowden (2000). All intensities are in units of $10^{-6} {\rm cts} \enspace {\rm s}^{-1}\,{\rm arcmin}^{-2}$.}
  \label{jkerp-E1-6:result}
  \end{center}
\end{table}

The intensities found for {\it field A} and {\it B} differ by about 40\% which can be 
attributed to a variation
in intensities with Galactic latitude and/or longitude (see Tab. \ref{jkerp-E1-6:result}). This 
behavior is {\em not} expected with the two distant component model proposed 
by \cite*{jkerp-E1-6:kun00}. A more detailed study of the $l,b$--dependency will be presented 
in a forthcoming paper.

\section{Summary} 
\begin{itemize}
\item The Local Bubble appears as a one component X--ray plasma with a temperature of 
$T = 10^{5.9}$\,K and an intensity of $350 \cdot 10^{-6} {\rm cts \enspace s^{-1}\,arcmin^{-2}}$ 
in the 1/4--keV energy regime. Moreover, the contribution of the halo exceeds the local 
distribution by a factor of four. 
\item The remaining SXRB is compatible with the following: A one component halo plasma with a 
temperature of $T =10^{6.2}$\,K and varying intensity in both latitude and longitude. Note, that
it is not necessary to introduce {\em more} than one halo component. Furthermore, an 
extragalactic component with an intensity of $170 \cdot 10^{-6} {\rm cts \enspace s^{-1}\,arcmin^{-2}}$ 
is found, which is consistent with the value provided by \cite*{jkerp-E1-6:bar96}
\item The variation with Galactic latitude and longitude suggests the existence of a smoothly
distributed halo plasma which surrounds the Milky Way.
\end{itemize}

\begin{acknowledgements}
The authors like to thank the Deutsches Zentrum f\"ur Luft-- und Raumfahrt for financial
support under grant No. 50\,OR\,0103.
\end{acknowledgements}


\begin{thebibliography}{}

\bibitem[\protect\astroncite{Barber et~al.}{1996}]{jkerp-E1-6:bar96}
Barber, C. R., Roberts, T. P., Warwick, R. S.\ 1996, MNRAS 282, 157B

\bibitem[\protect\astroncite{Hartmann \& Burton}{1997}]{jkerp-E1-6:har97}
Hartmann, D., Burton, W. B.\ 1997, Atlas of galactic neutral hydrogen, Cambridge University Press

\bibitem[\protect\astroncite{Hasinger et~al.}{2001}]{jkerp-E1-6:has01}
Hasinger, G., Altieri, B., Arnaud, M., Barcons, X., Bergeron, J., Brunner, H., Dadina, M., Dennerl, K., Ferrando, P., Finoguenov, A., Griffiths, R. E., Hashimoto, Y., Jansen, F. A., Lumb, D. H., Mason, K. O., Mateos, S., McMahon, R. G., Miyaji, T., Paerels, F., Page, M. J., Ptak, A. F., Sasseen, T. P., Schartel, N., Szokoly, G. P., Trümper, J., Turner, M., Warwick, R. S., Watson, M. G.\ 2001, A\&A 365L, 45H 

\bibitem[\protect\astroncite{Kuntz \& Snowden}{2000}]{jkerp-E1-6:kun00}
Kuntz, K. D., Snowden, S. L.\ 2000, ApJ 543, 195K

\bibitem[\protect\astroncite{Pietz et~al.}{1998}]{jkerp-E1-6:pie98}
Pietz, J., Kerp, J., Kalberla, P. M. W., Burton, W. B., Hartmann, Dap, Mebold, U.\ 1998, A\&A 332, 55P

\bibitem[\protect\astroncite{Snowden et~al.}{1998}]{jkerp-E1-6:sno98}
Snowden, S. L., Egger, R., Finkbeiner, D. P., Freyberg, M. J., Plucinsky, P. P.\ 1998, ApJ 493, 715S

\bibitem[\protect\astroncite{Snowden et~al.}{1991}]{jkerp-E1-6:sno91}
Snowden, S. L., Plucinsky, P. P., McCammon, D., Freyberg, M. J., Schmitt, J. H. M. M., Tr\"umper, J.\ 1991, BAAS 23, 1400S

\end{thebibliography}
\end{document}